\documentclass[twocolumn,showpacs,preprintnumbers,amsmath,amssymb]{revtex4}

\usepackage{graphicx}
\usepackage{dcolumn}
\usepackage{bm}

\newcommand{\beq}[1]{  \begin{equation} \label{#1} }  
\newcommand{\eeq}{     \end{equation}}  
\newcommand{\bal}[1]{\begin{align} \label{#1} }

\newcommand{\rf}[1]{(\ref{#1})}

\def\bd#1{\mbox{\boldmath$\displaystyle\mathbf{#1}$} }
\def\dd{\operatorname{d}} 
\def\eps{\epsilon}

\begin{document} 

\title{The worm-like chain model at small and large stretch}  

\author{ Andrew N. Norris} 

  \email{norris@rutgers.edu}
\affiliation{Mechanical and Aerospace Engineering, Rutgers University, Piscataway NJ 08854}

\date{\today}

\begin{abstract}

The relation between force and stretch  in the worm-like chain model of entropic elasticity is examined.  Although no closed-form expression is valid for all values of forcing,  solutions in the form of asymptotic series can be obtained under  conditions of small and large applied force.   The small and large stretch limits correspond to regular and boundary layer perturbation problems, respectively.  The perturbation problems are solved and series solutions obtained for force as a function of stretch. 
The form of the asymptotic series suggest a uniform approximation valid for all stretch that is an improvement on existing  approximations.   

\end{abstract}

\pacs{87.15.-v, 46.15.Ff, 82.37.Rs, 87.16.Ac}

\maketitle


\section{Introduction}

The worm-like chain  (WLC) is a model of entropic elasticity \citep{Doi86} for a macromolecule under thermal agitation.  The main  feature of the  model, as compared to simpler ones such as the freely jointed chain (FJC) model \citep{Flory69},  is the inclusion of  bending energy.    Applications of the WLC model range from  macroscopic elasticity of rubber and elastomers \citep{Ogden06} to DNA unfolding \cite{Bustamante00}.  With the increase in interest and application there is a need to more clearly understand how the WLC model relates  mechanical parameters, and in particular, the  relation between the force applied at the chain ends and the stretch.   This is complicated by the implicit and complex functional dependence in the model.  

The objective of this paper is to provide,  for the first time, explicit analytical expressions for the applied force as a function of the stretch of the WLC.  We begin with a brief introduction of the WLC model, and a review of existing closed-form approximations to the force-stretch relationship. 

\section{The worm-like chain model}\label{sec1}
An excellent overview of the theory  underlying  the WLC model is given by  \citet{Marko95}.  
Consider a uni-dimensional flexible chain of total length $L_0$  with end-to-end applied force ${\bd F}$. The  free energy of the chain is 
\beq{83}
E_{WLC} = 
\int_0^{L_0} \dd l \, \big( \frac{L_p}{2\beta} |{\bd t}'|^2  - {\bd t}\cdot {\bd F}\big),  
\eeq
where 
$L_p$  is the persistence length,  ${\bd t}(l)$ is the unit tangent vector, and $\beta = (kT)^{-1}$.   The applied force results in average stretch $z$ at temperature $T$.  

\begin{figure}[b]
				\begin{center}	
				    \includegraphics[width=3.in , height=2.5in 					]{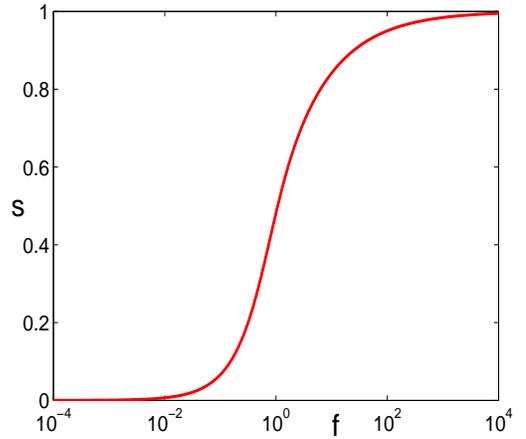} 
	\caption{The WLC relation between stretch $s$ and applied force $f$.  The numerical method is summarized in the Appendix. }
		\label{f1} \end{center}  
	\end{figure}

The natural non-dimensional units of force and  stretch are
\beq{033}
f = \beta L_p F, 
\qquad 
s =  {z}/{L_0}. 
\eeq
Using standard arguments from statistical mechanics \cite{Marko95,Bouchiat99} 
\beq{79}
 s =  \frac{L_p}{L_0} \frac{\partial \ln Z}{\partial f}, 
\eeq
where $Z$ is the partition function over all possible states.  it is certainly the case in elastomers, and generally true for DNA, that the persistence length  is much less than the unfolded molecule end-to-end length.   The large parameter  $ L_0/L_p \gg 1$ ensures that 
$\ln Z$, which  can be identified as chain entropy,   
 is dominated  by the lowest energy state,.  As a result  \citep{Marko95}
$Z \approx - (L_0/L_p) \eps_0$, where $\eps_0$ is a nondimensional energy, defined as 
\beq{80}
\eps_0 = \min\limits_{\psi} 
\int_{-1}^1 \dd x\, \big[\frac12 (1-x^2)(\psi')^2 - f x \psi^2\big] .  
\eeq
The probability density function  is normalized  $\langle \psi, \psi \rangle = 1$ with respect to the inner product
\beq{7}
\langle \psi ,\phi \rangle = \int_{-1}^1 \dd x\, \psi(x)\phi(x) . 
\eeq
The function $\psi$  is  smooth and  bounded for all $-1\le x\le 1$. 
The stretch is then 
\beq{82}
s = -\frac{\partial \eps_0}{\partial f} = \int_{-1}^1 \dd x \, x\psi^2 .
\eeq
The two terms in $\eps_0$ of \rf{80} correspond to the bending and work terms in the original energy $E_{WLC}$, and the specific form of the integrands is associated with  rotational invariance about the force axis, with ${\bd t}\cdot{\bd F} = F \cos \theta = Fx$.

The WLC problem therefore requires finding stationary values of  the functional  
\bal{84}
\Gamma (\psi) =& 
\int_{-1}^1 \dd x\,  \frac12 (1-x^2)(\psi')^2
 - f \big( \int_{-1}^1 \dd x\,  x\psi^2  -s \big) 
 \nonumber \\ & \qquad  -
 \eps_0  \big( \int_{-1}^1 \dd x\,  \psi^2  -1 \big) . 
\end{align}
$\Gamma (\psi) $ contains the bending energy term  plus two constraints involving the first two moments of the  function $\psi$.   The normalization 
$\langle \psi, \psi \rangle = 1$  defines $\psi$ as a  probability density function, while the 
constraint $\rf{82}_2$ defines the stretch $s$. 
We may consider the stretch  as given, so that $f$ and $\eps_0$ are   Lagrange multipliers, and the Euler-Lagrange equation  is
\beq{85}
\frac12 [ (1-x^2)\psi']' + fx \psi + \eps_0 \psi = 0, \quad -1\le x\le 1,
\eeq
  The objective is  to find the lowest value of $\eps_0$, and  the force $f$ is then uniquely determined as a function of $s$.  This dictates an  indirect  procedure: consider $f$ as given, and find $\eps_0$, the lowest eigenvalue of the differential operator that depends upon $f$.  Then $s$ is determined as a function of $f$ via either  formulas given by eq. \rf{82}.
 Note that the value of $\Gamma $ at the minimum is $\gamma_0 = \eps_0  + sf $, which is the Legendre transform of $\eps_0$ with  $f =  \partial \gamma_0 /\partial s$.  
  The 2D version of  eq. \rf{85}  reduces to the  Mathieu differential equation with solution in terms of Mathieu functions \cite{Prasad05}.   Prasad et al. \cite{Prasad05}  derived small and large force limits for the WLC in two dimensions using this approach. The focus here is on the 3D problem only. 

Figure \ref{f1} shows the characteristic WLC curve, obtained from eqs. \rf{82} and \rf{85} using a numerical   method based on \cite{Marko95}, see the Appendix.  There are other ways to find $f=f(s)$, e.g. by solving the ODE using a shooting method  \cite{Bouchiat99}.  The important issue  is not, however, the numerical determination of the curve, but finding a suitable analytic approximation.   An excellent first step in this direction was made by Marko-Siggia \cite{Marko95}  who showed the leading order behavior for $f\ll 1$ and for $f\gg 1$ is $f=\frac32 s$  and $f^{-1} = 4(1-s)^2$, respectively.  Motivated by this limiting behavior  they 
 suggested the  approximate functional form
\beq{91}
f_{MS} = \frac1{4(1-s)^2} - \frac14 + s. 
\eeq
 This simple formula reproduces the small and large stretch leading order response in the respective limits. 
 Ogden et al. \cite{Ogden07} examined several alternative approximations based on intelligent curve fitting to the $f-s$ data in \cite{Bouchiat99}. The simplest formula, which they called  $WLC_3$, is just the Marko-Siggia approximation with a single  term added:
\beq{011}
WLC_3 =  \frac1{4(1-s)^2} - \frac14 + s-\frac34 s^2.
\eeq
The extra quadratic term  $-\frac34 s^2$ produces a dramatic improvement, see Fig. \ref{f4}. The root mean square  error of $WLC_3$ is $0.013$ as compared with $0.339$ for $f_{MS}$. 
 The analytical results of this paper will help explain this roughly 25-fold increases in accuracy. We will return to consider $WLC_3$ in Section \ref{sec4}  after deriving the small and large stretch approximations.  The principal results of the papers are summarized next.

\subsection{Summary of the main results}
\begin{figure*}[htbp]
\centering
\begin{tabular}{cc}
 \includegraphics[width=3.in , height=2.5in 					]{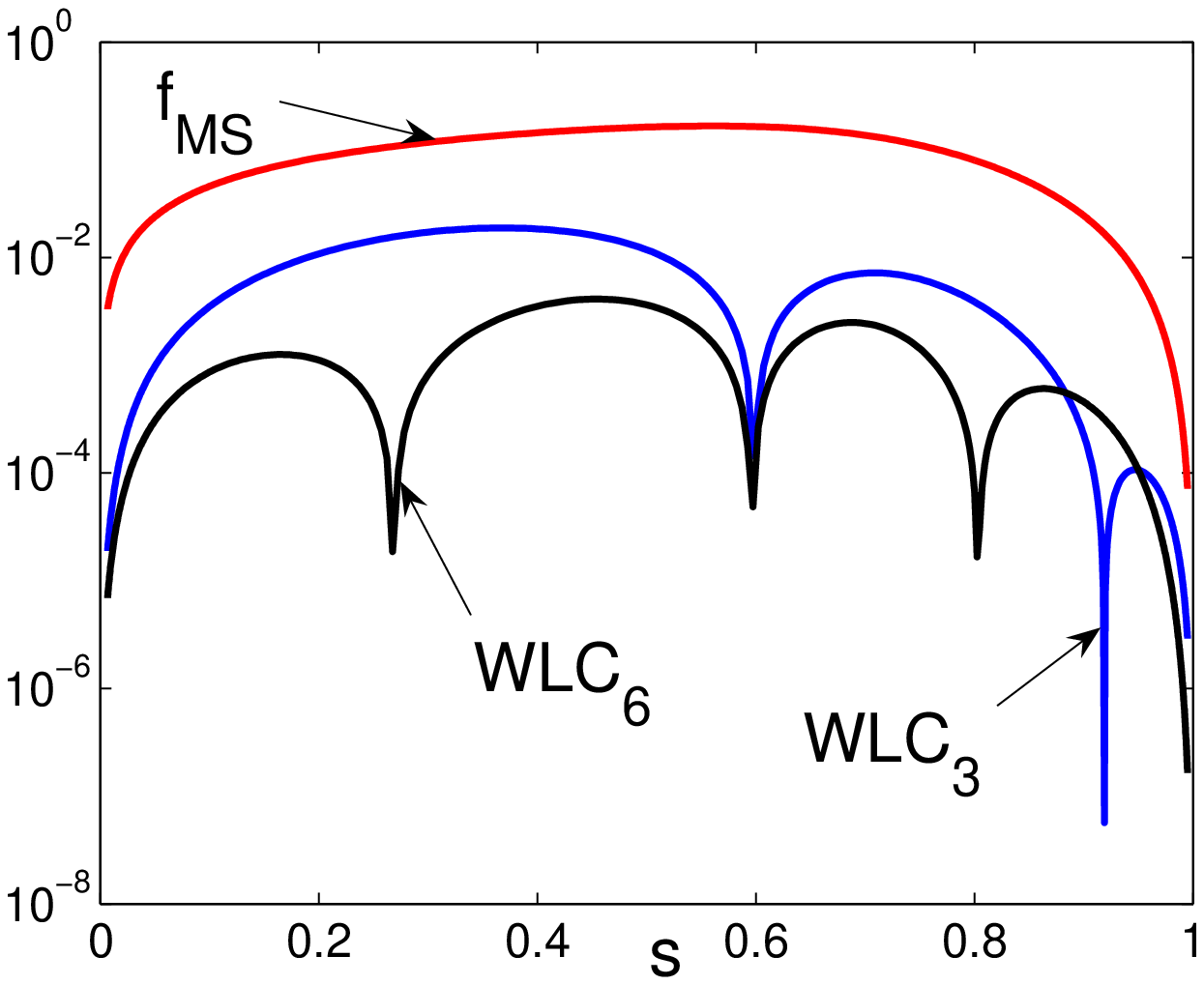} 
 &
 \includegraphics[width=3.in , height=2.5in 					]{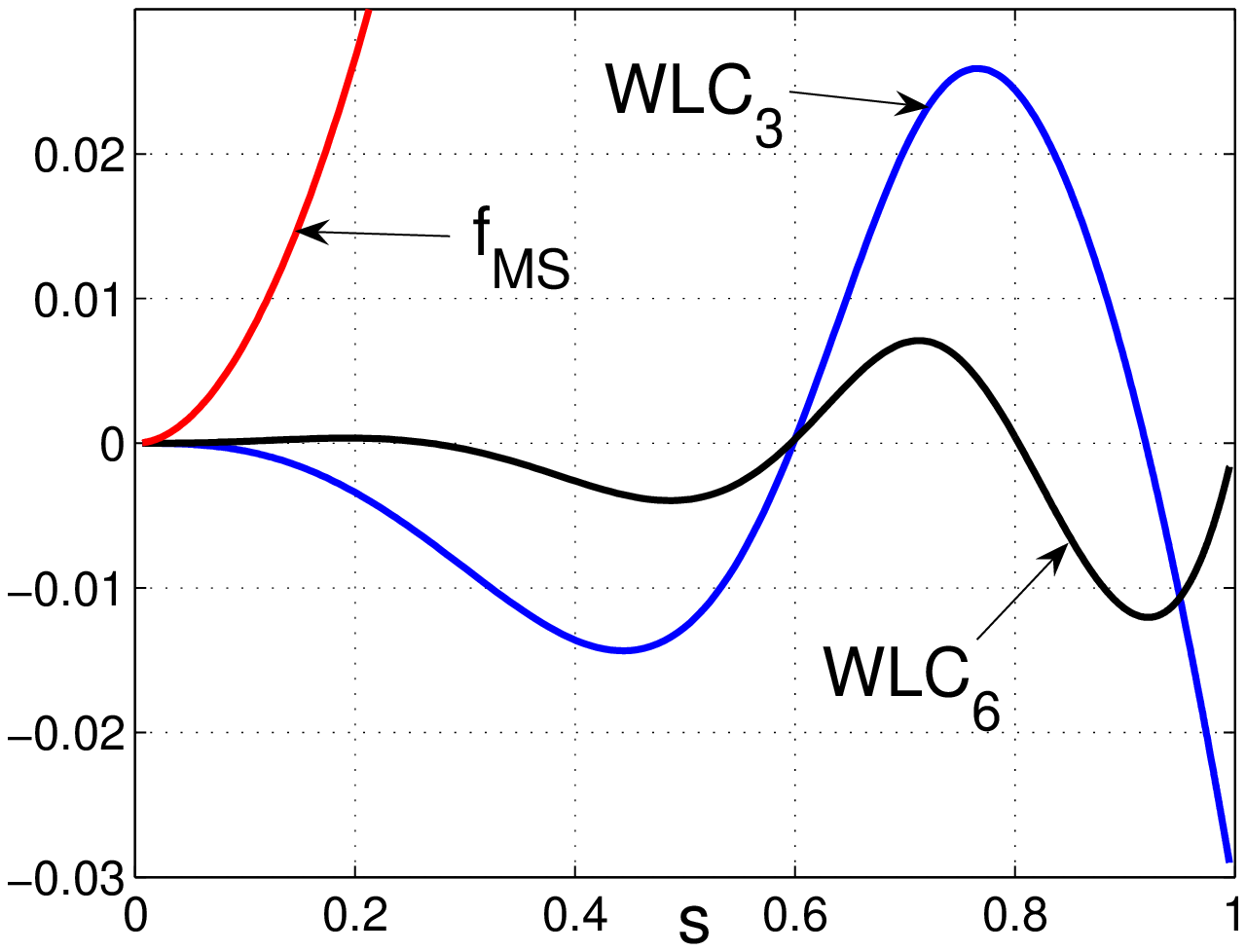} 
 \\
(a) & (b)\\
\end{tabular}
	\caption{ The approximants $f_{MS}$, $WLC_3$ and $WLC_6$ of eqs. \rf{91}, \rf{011} and \rf{013} compared.   The curves show the relative  error compared with the exact solution on a log scale (a) and absolute value (b).} 
		\label{f4}  
\end{figure*}

The small and large stretch expansions are 
\begin{eqnarray}\label{891}
f =  \left\{ 
\begin{array}{l}
\frac{3}{2} s  + \frac{33}{20} s^3 
+ \frac{3393}{1400} s^5 + \ldots ,   
\\
  \\ 
\frac1{4(1-s)^2} +\frac1{32} + \frac{3}{64} (1-s)  + \frac{2559}{32768} (1-s)^2 
+ \ldots  , 
\end{array} 
\right.
\end{eqnarray}
valid for $s\ll 1$ and $1-s\ll 1 $, respectively.  Based on these limiting forms, and some numerical experimentation, we find that the following approximation to $f$  shows significant  improvement on $WLC_3$, 
\beq{013}
WLC_6 =  \frac1{4(1-s)^2} - \frac14 + s-\frac34 s^2
+ \frac{1}{64}s^3 (3 - 5s)(19 - 20s) .
\eeq
This has rms error of $0.0047$ and is compared with $WLC_3$ in Fig. \ref{f4}.

The remainder of the paper is organized as follows.  The asymptotic series of eq. \rf{891} are derived in Sections \ref{sec2} and \ref{sec3}.  The small stretch regime is considered first in Section \ref{sec2}, where the solution is developed using regular perturbation methods.   Large stretch is examined in Section \ref{sec3}.  Although the problem is a singular perturbation, it is reduced to a regular perturbation expansion using an inner scaled variable.      The two asymptotic series are compared with the exact solution in Section \ref{sec4}.  The new and improved approximate formula valid for all values of stretch, large and small,  is proposed after some numerical experimentation.

\section{Small stretch expansion}\label{sec2}

\subsection{Perturbation theory}

Under small stretch, or equivalently small applied force, the WLC  equation reduces to a regular perturbation problem.  
Define 
\beq{1}
L = \frac{\dd}{\dd x} (1-x^2) \frac{\dd}{\dd x}, 
\eeq
then with the replacements $\eps_0 \rightarrow \frac12 \lambda$ and $f\rightarrow \frac12 \eps$ the equation \rf{85} becomes
\beq{2}
L \psi  + \lambda \psi + \epsilon x \psi = 0, \quad -1\le x \le 1. 
\eeq
The small stretch limit corresponds to $\eps \ll 1$.  We seek solutions to eq. \rf{2} in the form of  
a regular perturbation expansion   
\begin{subequations}
\bal{3a}
\psi  &= \psi_0 + \epsilon   \psi_1 + \epsilon^2   \psi_2 + \ldots ,  
\\
\lambda  &= \lambda_0 + \epsilon   \lambda_1 + \epsilon^2   \lambda_2 + \ldots .   
\end{align}
\end{subequations}
Substituting these  into eq. \rf{2} and identifying terms of like order in the perturbation parameter $\eps$ yields a sequence of equations. 
The first few  of order $\eps^0$, $\eps^1$ and $\eps^2$, are  respectively, 
\begin{subequations}
\bal{4a}
L_0\psi_0  &= 0,  
\\
L_0\psi_1+ x\psi_0 + \lambda_1 \psi_0  &= 0,  
\\   
L_0\psi_2+ x\psi_1 + \lambda_1 \psi_1 + \lambda_2 \psi_0  &= 0.   
\end{align}
\end{subequations}
where 
\beq{5}
L_0 \equiv    L + \lambda_0. 
\eeq
Although the WLC corresponds to $\lambda_0 = 0$, it is useful to first consider the perturbation of an arbitrary ground state.   

The form of the O$(\eps^k)$, $k\ge 1$, equation is 
\beq{6}
L_0\psi_k+ x\psi_{k-1} + \lambda_1 \psi_{k-1} + \lambda_2 \psi_{k-2} +\ldots + \lambda_k \psi_0  = 0.
\eeq
The unperturbed solution $\psi_0 (x)$ is either an even or an odd function of $x$.   It follows that 
$\psi_k$ has the same or opposite parity depending as $k$ is even or odd, respectively. 
We assume the unperturbed solution  is normalized $ \langle \psi_0,\psi_0\rangle  =1$.

The operator $L_0$ is self adjoint with respect to the inner product \rf{7},   
   implying the solvability condition at 
O$(\eps^k)$ is 
\beq{8}
\lambda_k 
+ \lambda_{k-1} \langle \psi_1,\psi_0\rangle 
+ \ldots 
+ \lambda_1 \langle \psi_{k-1},\psi_0\rangle 
+ \langle x\psi_{k-1},\psi_0\rangle 
= 0. \nonumber 
\eeq
The solvability condition essentially ensures that the  solution to eq. \rf{6} can be expressed in terms of a sum of Legendre polynomials that are regular at the end points, i.e. $P_n$.  However, the expression for $\psi_k$ has no component corresponding to  $\psi_0$, in other words, $\langle \psi_k,\psi_0\rangle = \delta_{k0} $.   Taking into account the parity of the successive terms  
 gives the succinct result
 \beq{102}
 \lambda_{2k-1} = 0, \quad \lambda_{2k} = - 
 \langle \psi_{2k-1},\psi_0\rangle  ,\quad k=1,2,\ldots .
 \eeq

Note that the  first few equations simplify to 
\begin{subequations}\label{-4}
\bal{-4a}
L_0\psi_0  &= 0,  
\\
L_0\psi_1+ x\psi_0   &= 0,  
\\   
L_0\psi_2+ x\psi_1  + \lambda_2 \psi_0  &= 0,
 \\   
L_0\psi_3+ x\psi_2  + \lambda_2 \psi_1  &= 0,
\\
L_0\psi_4+ x\psi_3  + \lambda_2 \psi_2  + \lambda_4 \psi_0 &= 0,
\\
L_0\psi_5+ x\psi_4  + \lambda_2 \psi_3  + \lambda_4 \psi_1 &= 0.
\end{align}
\end{subequations}
We will solve these for the WLC problem, which corresponds to the lowest eigenvalue. Before considering the WLC specifically,  we note some properties of the    eigenvalue perturbation that are valid for any eigenvalue. 

\subsection{$\lambda_2$ for any initial state}
The unperturbed eigenvalue problem is Legendre's equation, and hence the most general form of the unperturbed solution is  
\beq{20}
\psi_0(x) = c_n P_n(x), \qquad \lambda_0 = n(n+1), 
\eeq
where $P_n$ is the  Legendre polynomial of order $n$ and the normalization factor is $c_n = \sqrt{n+\tfrac12}$. 

Using the identity \citep{Abramowitz74}
\beq{21}
(2k+1)xP_k =  k P_{k-1} + (k+1)P_{k+1}, 
\eeq
it is easy to show that 
\beq{22} 
L_0  (    P_{k+1} -P_{k-1}(1-\delta_{k0}) ) + 2(2k+1) x P_k = 0. 
\eeq
Hence, the first correction to the unperturbed mode is 
\beq{23}
\psi_1 =  
\frac{c_n}{2(2n+1)}\, 
(    P_{n+1} -P_{n-1}(1-\delta_{n0}) ). 
\eeq
The first correction to the eigenvalue follows from the  identities \citep{Abramowitz74}
\beq{24}
\langle xP_l,P_n\rangle 
= \begin{cases}
\frac{2n}{(2n-1)(2n+1)}, & l=n-1, \\
\frac{2(n+1)}{(2n+1)(2n+3)}, & l=n+1,
\end{cases}
\eeq
as
\beq{25}
\lambda_2 = [2(2n-1)(2n+3)]^{-1}. 
\eeq
Note that $\lambda_2 >0$ for all $n$ except $n=0$, which has the lowest eigenvalue.   We now consider the lowest energy state specifically and continue the perturbation expansion to higher orders. 

\subsection{The lowest eigenvalue}
We focus on the unperturbed solution for  $n=0$, which has the lowest initial energy.  The analysis of the previous subsection gives the first two terms in the eigenvalue and eigenfunction expansions as  $\lambda_0 = 0$, $\lambda_2 = - \frac16$,  and 
$\psi_0 = c_0P_0$, $\psi_1 = \frac{c_0}{2} P_1$,  with 
 $c_0 = 1/\sqrt{2}$.   These are the solutions of the first two in the hierarchy of equations \rf{-4}.    The next two are then solved  to obtain $\psi_2$ and $\psi_3$, from which 
the next term in the eigenvalue   expansion, $\lambda_4$,  follows from eq. \rf{102}. 

In this manner  the first six equations given in \rf{-4} may be solved successively.  The  terms in the eigenfunction expansion were obtained using Mathematica, 
\begin{subequations}
\bal{36}
\psi_0 & = c_0P_0, \quad \psi_1 = \tfrac{1}{2} c_0P_1, \quad  \psi_2 = \frac{c_0}{18}  P_2, 
\\
\psi_3 &= \frac{c_0}{3.4.5.6}  (P_3 -11 P_1),
\\
\psi_4 &= \frac{c_0}{7.8.9.10}  (\tfrac25  P_4 -\tfrac{215}{9}P_2),
\\
\psi_5 &= \frac{c_0}{2^73^45^2}  (\tfrac8{21}  P_5 -\tfrac{212}{3}P_3+\tfrac{7520}{7}P_1),
\end{align}
\end{subequations}
and the corresponding expansion of the eigenvalue is 
\beq{09}
\lambda = -\frac16 \eps^2 + \frac{11}{1080}\eps^4  -\frac{47}{34020} \eps^6 + \text{O}(\eps^8).
\eeq
The procedure can be continued; however the coefficients quickly become more unsightly.   

\subsection{Small stretch expansion}
Taking into account the factor of $1/2$ difference between eq. \rf{2} and the WLC equation \rf{85}, the above analysis implies that the lowest perturbed  energy  is 
\beq{87}
\eps_0 = -\frac{1}{3} f^2 + \frac{11}{5.27} f^4 
- \frac{8.47}{5.7.9.27} f^6 + \ldots . 
\eeq 
The stretch follows from eq. \rf{82},
\beq{88}
s =  \frac{2}{3} f  - \frac{44}{5.27} f^3 
+ \frac{16.47}{5.7.9.9} f^5 + \ldots , 
\eeq 
and inverting the series  gives %
\beq{89}
f =  \frac{3}{2} s  + \frac{33}{20} s^3 
+ \frac{9.13.29}{1400} s^5 + \ldots .
\eeq 

\begin{figure*}
\centering
\begin{tabular}{cc}
 \includegraphics[width=3.in , height=2.5in 					]{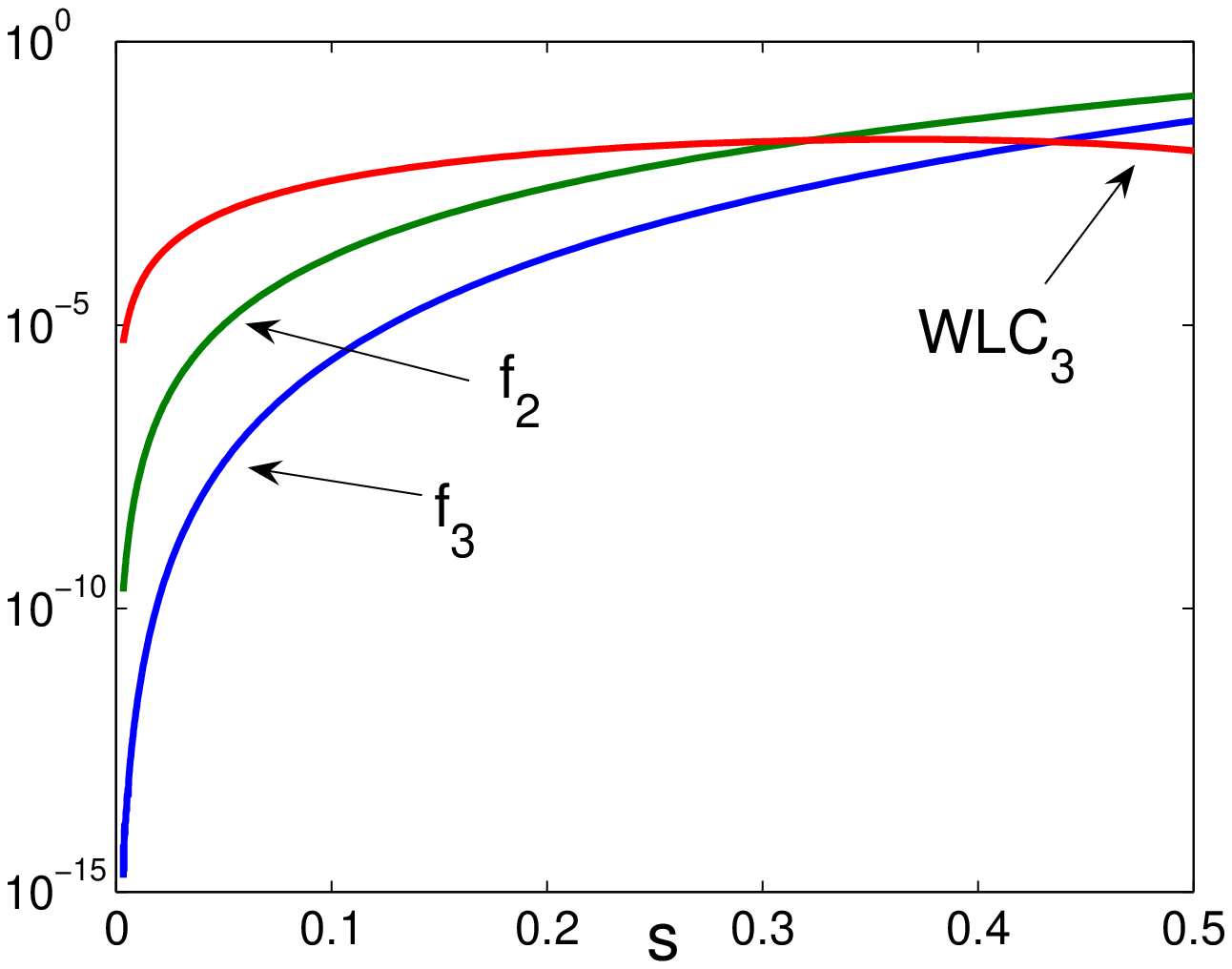} 
 &
 \includegraphics[width=3.in , height=2.5in 					]{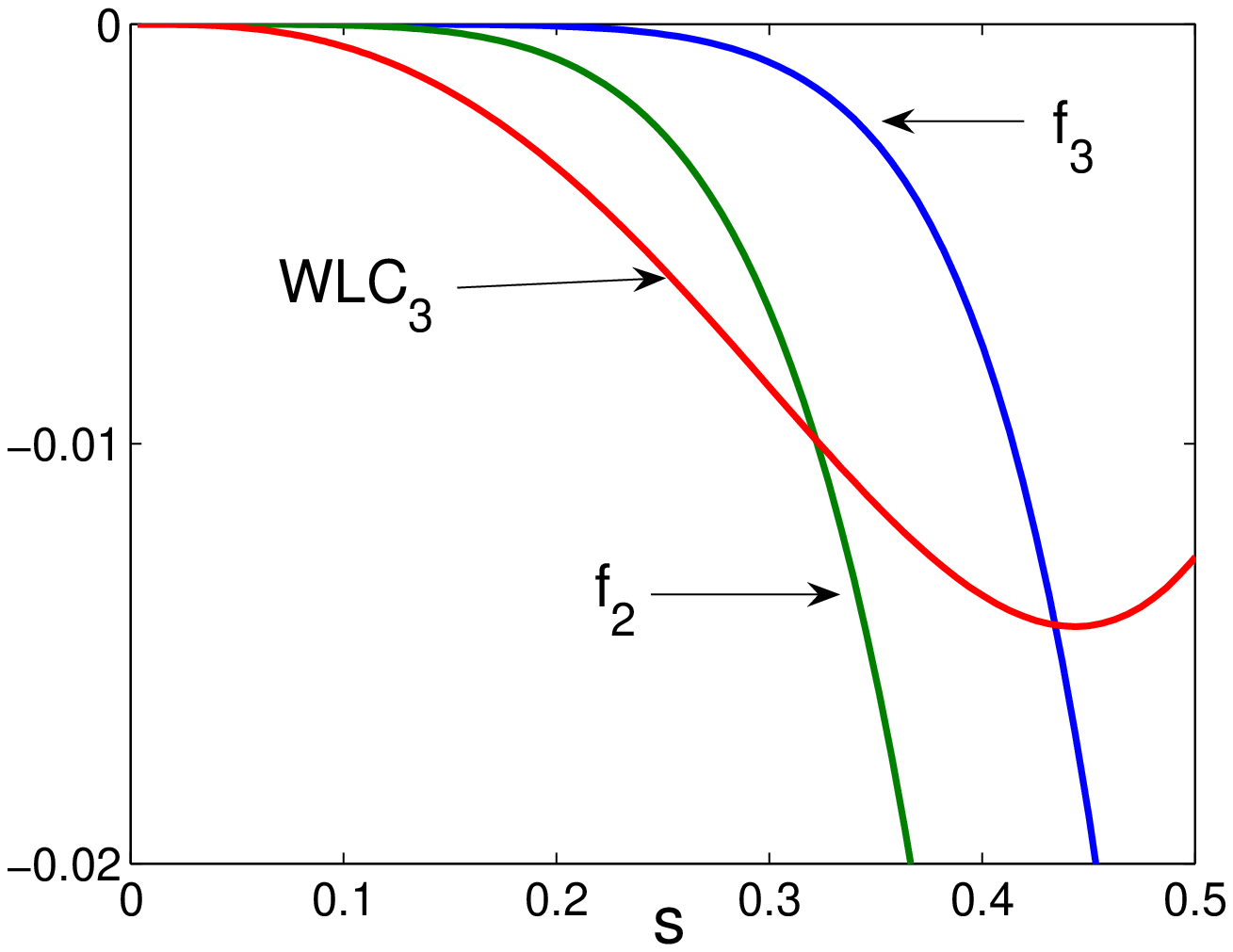} 
 \\
(a) & (b)\\
\end{tabular}
\caption{ The relative  error of the perturbation solution compared with the exact solution at small stretch on a log scale (a) and absolute value (b). $f_3(s)$ is the three term  expansion of  eq. \rf{89}, and $f_2(s)$  is the first   two terms  only.  The relative error of the $WLC_3$ approximation of eq. \rf{011} is also shown. }
		\label{f2}
\end{figure*}

The accuracy of the small stretch  expansion is shown in Fig. \ref{f2}, with $WLC_3$ used as a comparison.  The relative error of the three term asymptotic series is less than $10^{-3}$ for $0\le s < 0.3$, but the approximation deteriorates at higher values, as expected. 

\section{Large stretch: a boundary \protect\\  layer approximation}\label{sec3}

\subsection{A singular perturbation problem}
The large stretch limit corresponds to large values of the applied force $f$ in eq. \rf{85}.  We therefore consider 
\beq{+2}
\tfrac12 L \psi  + \lambda \psi + \eps^{-2} x \psi = 0, \quad -1\le x \le 1, 
\eeq
for $\eps \ll 1$.  The second order differential operator $L$ is defined in eq. \rf{1}, and the factor of $1/2$ is introduced for convenience. 
Equation \rf{+2} defines a singular  perturbation problem  for $\psi(x)$, describing a 
boundary layer solution that is non-zero only near $x=1$.
  In order to deduce this   introduce the boundary layer variable  
\beq{+3}
X = (1-x)\eps^{-1}.
\eeq
Let $\Psi (X) = \psi(x)$, and  define $\Lambda$ by 
\beq{+4}
\lambda = -\eps^{-2} +\eps^{-1}\Lambda, 
\eeq
then the equation for $\Psi$ becomes
\beq{+5}
(X\Psi ')' + (\Lambda -X)\Psi - (\eps /2)\, (X^2\Psi ')' = 0, 
\eeq
for $0\le X \le  {2}/{\eps}$. This is now a regular perturbation problem in terms of the rescaled inner coordinate $X$.  Note that  the range of $X$ depends upon the small parameter $\eps$, although this is not a serious  complication since the effective range of $X$ is the positive real axis. 

Assuming the regular perturbation expansion   
\begin{subequations}
\bal{+3a}
\Psi  &= \Psi_0 + \epsilon   \Psi_1 + \epsilon^2   \Psi_2 + \ldots ,  
\\
\Lambda  &= \Lambda_0 + \epsilon   \Lambda_1 + \epsilon^2   \Lambda_2 + \ldots ,   
\end{align}
\end{subequations}
gives the sequence of equations
\begin{subequations}
\bal{+4a}
(X\Psi_0 ')' + (\Lambda_0 -X)\Psi_0   &= 0,  
\\
(X\Psi_1 ')' + (\Lambda_0 -X)\Psi_1  + \Lambda_1\Psi_0  -\tfrac12  (X^2\Psi_0 ')' &= 0,    
\end{align}
\end{subequations}
etc. 
The solution of the first equation, of order $\eps^0$, 
is 
\beq{+51}
\Psi_0 (X) = C_0 e^{-X}, 
\qquad 
\Lambda_0 = 1, 
\eeq
where normalization implies $C_0 = \sqrt{2}$. The next  equation, 
of order $\eps^1$, becomes  
\beq{+6}
(X\Psi_1 ')' + (1 -X)\Psi_1  + \big( \Lambda_1   +X -\tfrac{1}{2}X^2 \big) \Psi_0 = 0. 
\eeq
The solvability condition  
\beq{+7} 
  \int_0^\infty \dd X (\Lambda_1 +X -\tfrac{1}{2}X^2 ) \Psi^2 (X) =0,
\eeq
implies the first correction  is $\Lambda_1 =  -\frac14$. 

It is evident that the  solutions have the form of the fundamental exponentially decaying solution $\Psi_0 (X)$ multiplied by polynomials in $X$. 
This suggests scaling $\Psi$ with respect to the leading order  solution, 
\beq{044}
\Psi (X) = g(X)\Psi_0 (X) .
\eeq
The equation for $g$ is 
\beq{0534}
Jg + \eps Hg + (\Lambda - 1 - \Lambda_1 \eps ) g = 0, 
\eeq
where the differential operators $J$ and $H$ are
\begin{subequations}
\bal{-77}
Jg(X) &= Xg''+(1-2X)g', 
\\
Hg (X)&=  
\big(     X- \frac{X^2}{2} -\frac1{4}\big)g + (X^2-X)g' - \frac{X^2}{2} g'' . 
\end{align}
\end{subequations}
Assuming the expansion
\beq{05677}
 g  =  g_0 + \epsilon    g_1 + \epsilon^2    g_2 + \ldots ,
\eeq
then $g_0=1$ and the equations for $g_1$ through  $g_4$ are 
\begin{subequations}\label{+34}
\bal{+34a}
&Jg_1     +  Hg_0 = 0,  
\\
&Jg_2   +  Hg_1  + \Lambda_2  = 0,  \label{+34b}
\\
&Jg_3   +  Hg_2  + \Lambda_2 g_1 + \Lambda_3   = 0,  
\\
&Jg_4   +  Hg_3  + \Lambda_2 g_2 + \Lambda_3 g_1 + \Lambda_4  = 0 , 
\end{align}
\end{subequations}
The procedure is then 
to find $g_1$ as the particular solution to eq. \rf{+34a}  and $\Lambda_2$   
follows from the solvability condition for eq. \rf{+34b}:
\beq{+78} 
  \int_0^\infty \dd X (\Lambda_2 +Hg_1 ) \Psi_0^2 (X) =0.
\eeq
These steps are repeated to find the successive functions $g_k$ and the eigenvalue coefficients
$\Lambda_k$. 

Equations  \rf{+34} were solved using Mathematica.  We omit the detailed form of the $g_k$ functions and focus on the eigenvalue solution which is all that is required for the WLC model,
\beq{009}
\lambda = -\frac{1}{\eps^2} +\frac{1}{\eps} -\frac1{4} -\frac1{64} \eps -\frac3{512} \eps^2 
-\frac{885}{262144} \eps^3
+ \text{O}(\eps^4).
\eeq

\begin{figure*}[htbp]
\centering
\begin{tabular}{cc}
 \includegraphics[width=3.in , height=2.5in 					]{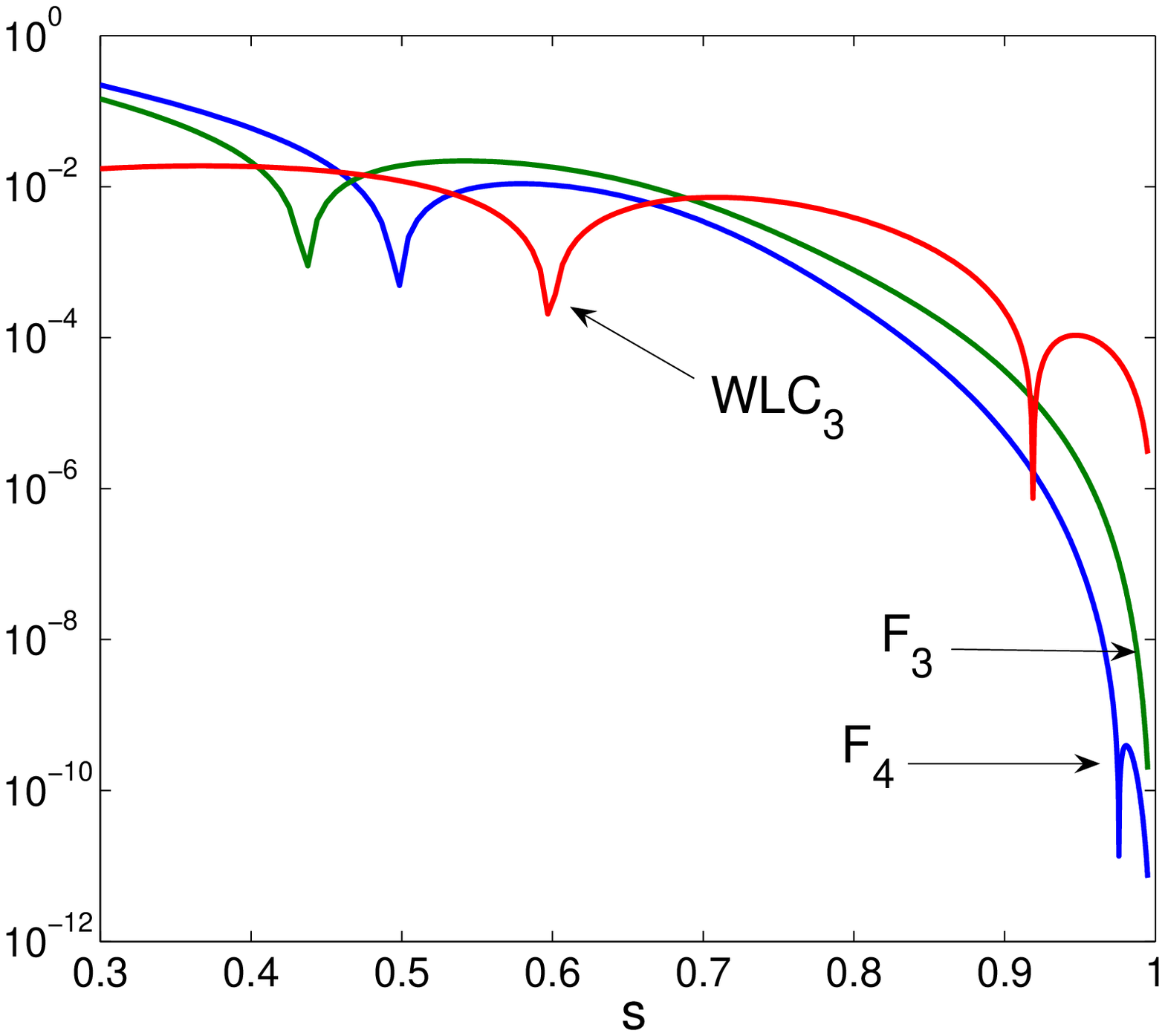} 
 &
 \includegraphics[width=3.in , height=2.5in 					]{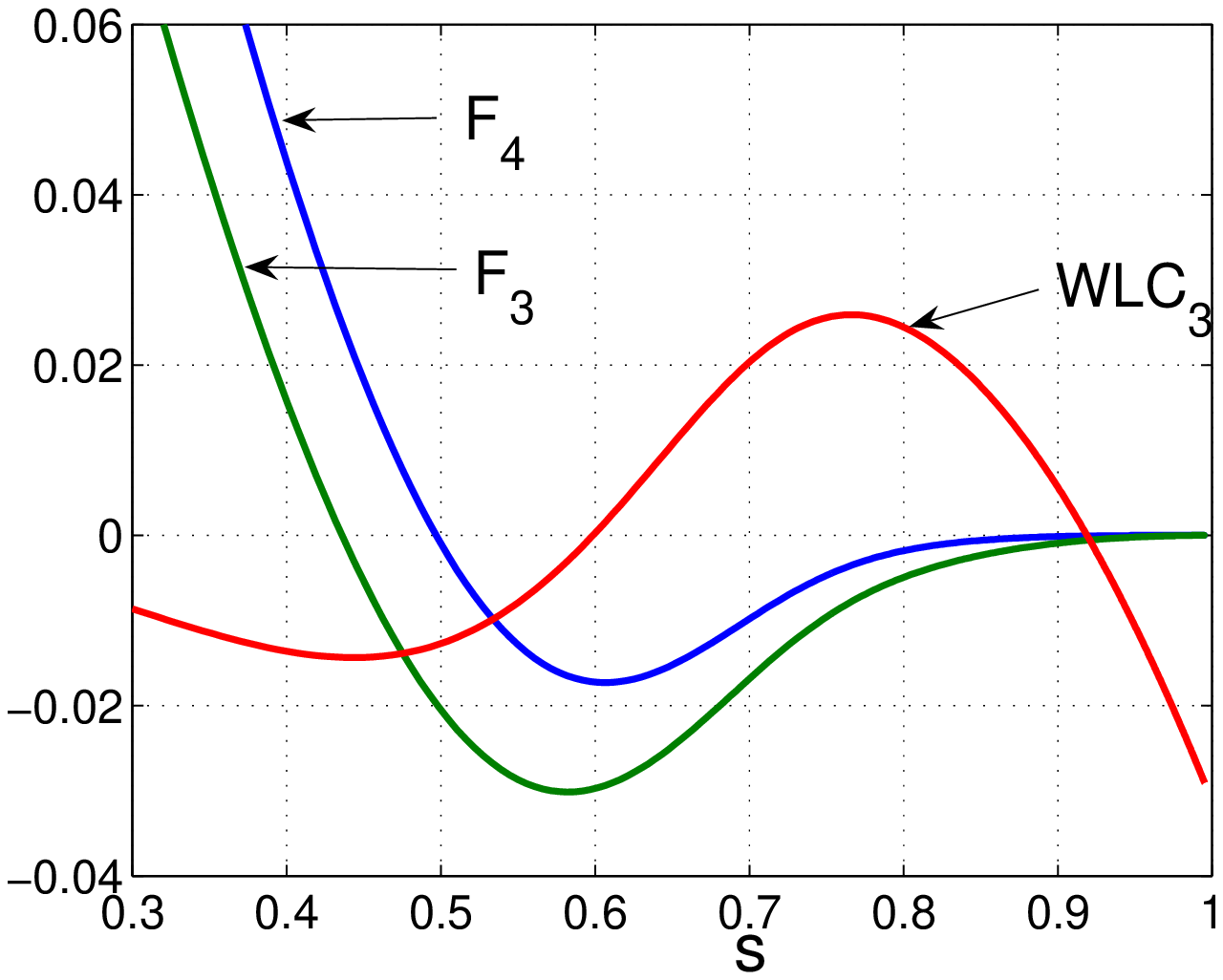} 
 \\
(a) & (b)\\
\end{tabular}
\caption{ The relative  error of the perturbation solution compared with the exact solution at large stretch on a log scale (a) and absolute value (b). $F_4(s)$ is the four term  expansion of  eq. \rf{-89}, and $F_3(s)$  is the first   three terms  only.   The relative error of the $WLC_3$ approximation of eq. \rf{011} is also shown. }
		\label{f3}
\end{figure*}

\subsection{Large stretch expansion}

The boundary layer solution  with $\eps = f^{-1/2}$ implies that the lowest energy state of the WLC has the large force expansion
\beq{-87}
\eps_0 = -f +f^{\frac12} -\frac1{4} -\frac1{64f^{\frac12}}   -\frac3{512f}  
-\frac{885}{262144  f^{\frac32}}
 + \ldots . 
\eeq 
The stretch is then 
\beq{-88}
s =  1 -\frac{1}{2f^{\frac12}}  -\frac1{2^7f^{\frac32}} -\frac3{2^9 f^2} -\frac{5.9.59}{2^{19} f^{\frac52}}
 + \ldots .  
\eeq 
Inverting the asymptotic series gives the desired expression for $f$ as a function of $s$,  
\beq{-89}
f =  \frac1{4(1-s)^2} +\frac1{32} + \frac{3}{64} (1-s)  + \frac{2559}{32768} (1-s)^2 
+ \ldots . 
\eeq 

The large  stretch  asymptotic expansion is illustrated  in Fig. \ref{f3}, with $WLC_3$ again  used as a comparison.  The relative error of the four  term  series is less than $10^{-3}$ for $0.8< s \le 1$, roughly.


\section{Numerical experiments }\label{sec4}

Comparison of the accuracy of the small and large stretch expansions in  Figs. \ref{f2} and \ref{f3}
indicate the at series as developed here are accurate to within on part in $10^3$ for the range $0\le s<0.3$ and 
   $0.8< s\le1$, with zero error at $s=0$ and $s=1$.  This Section  examines the question of finding an approximation that is uniformly valid over the entire range of stretch.

The difference between the exact force function and $WLC_3$ at small stretch follows from 
eqs.  \rf{011} and \rf{891} as
\begin{eqnarray}\label{92}
f - WLC_3 =  \left\{ 
\begin{array}{l}
 \frac{s^3}{8}\big(\frac{26}{5} - 10 s + \frac{1293}{175}s^2 \big)  + \text{O}(s^6) ,   
\\
  \\ \frac1{32}   -\frac{29}{64} (1-s)  + \frac{27135}{32768} (1-s)^2 
  \\ \qquad  \qquad  \qquad  \qquad 
+  \text{O}\big( (1-s)^3\big) . 
\end{array} 
\right.
\end{eqnarray}

The  term $ -\frac34 s^2$ that distinguishes $WLC_3$ from the Marko-Siggia approximation \rf{91}
therefore exactly cancels the error in the latter at O$(s^2)$ in the small stretch limit.  

The quadratic for large stretch has zeros at $s=0.9191$ and $s=0.5337$.  The first zero, being close to $s=1$ can be attributed as the cause of the zero of $f-WLC_3$ at $s\approx 0.9189$, see Fig. \rf{f4}.  The zeros of the quadratic in \rf{92} for small stretch are complex.  However, as Fig. \rf{f4} indicates   $f-WLC_3$  has a second zero at  $s\approx 0.5986$. 
This property of $WLC_3$, that it is exact at $s\approx 0.6$ and $s\approx 0.92$, partly explains its success as a uniform approximant.  This suggests that any attempt at improving on $WLC_3$ should maintain these zero crossings, and preferably increase the number of zero crossings.  

At the same time wish to improve the accuracy at large stretch, requiring that the new approximation, say $f^*$, is exact at $s=1$.  Consider the two parameter extension 
$f^* = WLC_3 + cs^3(a-s) $, then the constraint  $f^*(1)=1/32$ implies $c = (a-1)/32$.  
Numerical experiments show that $f^* = WLC_3 + \frac{s^3}{32} \frac{(a-s)}{(a-1)}$
 is not an improvement on $WLC_3$ no matter what value of $a$ is chosen.  We therefore consider the two-parameter function 
 \beq{0--}
f^* = WLC_3 + \frac{s^3}{32} \frac{(a-s)(b-s)}{(a-1)(b-1)}. 
\eeq
Using \emph{fminsearch} in Matlab to minimize the root mean square error $\langle f-f^*,f-f^* \rangle^{1/2}$ gives $a=0.5986$, $b=    0.9458$.  Surprisingly, the value of $a$ is precisely  (to  within four significant figures)  the existing zero crossing of $WLC_3$.   In order to provide an approximation that is not too difficult to remember, we suggest rounding $a$ and $b$ up to $0.6$ and $0.95$, respectively. We call the resulting approximant $WLC_6$, 
\beq{-013}
WLC_6 =  \frac1{4(1-s)^2} - \frac14 + s-\frac34 s^2
+ \frac{100}{64}s^3 (0.6 - s)(0.95 - s) .
\eeq
The rms error incured by $WLC_6$ is $0.0047$, as compared with $0.0045$ for $f^*$ of \rf{0--} with 
$a=0.5986$, $b=    0.9458$.   The rms errors for  $f_{MS}$ and $WLC_3$ are $0.3386$ and  $0.0132$, 
respectively.  These numbers indicate the remarkable accuracy of all three approximations to the exact force function  $f(s)$.

\appendix  
\section{Exact solution}

\begin{figure}
				\begin{center}	
				    \includegraphics[width=3.in , height=2.5in 					]{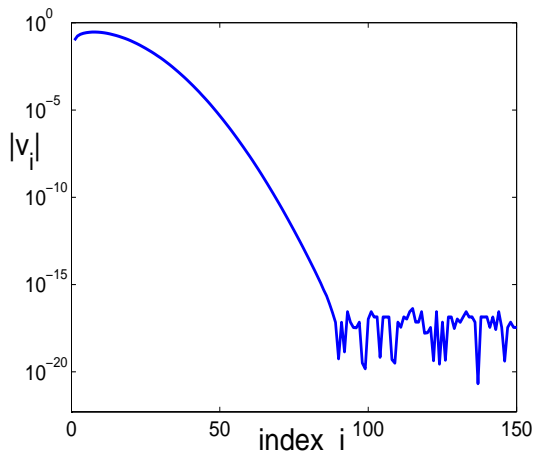} 
	\caption{The  eigenvector components for $f= 10^4$ solved using Matlab with $N=300$.  Only the first 150 components are displayed, the rest are in the noise floor.   }
		\label{f5} \end{center}  
	\end{figure}

The exact solution for $s = s(f)$ can be determined  numerically quite easily \citep{Marko95}.
Define  two symmetric matrices of size $(N+1)\times (N+1)$ with elements  
\beq{-11}
D_{ij} = \frac{i(i+1)}{2}\delta_{ij}  ,
\qquad 
S_{ij} =  \frac{ i \delta_{i-1,j} + j \delta_{i,j-1} }{\sqrt{ (2i+1)(2j+1)}},  
\eeq
for $i,j = 0,1,2,\ldots, N$. 
Then for a given $f$, determine the minimum eigenvalue of ${\bd D} - f {\bd S}$ and its eigenvector  ${\bd v}$. The strain is then 
\beq{-12}
s = { {\bd v}^t {\bd S}{\bd v} }/ ( {\bd v}^t  {\bd v}) . 
\eeq
This algorithm can be effectively implemented  in Matlab by using sparse matrix methods and the Matlab function \emph{eigs} to find the single lowest eigenvalue.  This is always negative but it is not always the smallest in magnitude, which is the criterion used in the function \emph{eigs}. This can be  circumvented by adding a multiple of the identity to ${\bd D} - f {\bd S}$ so that the lowest eigenvalue is also the smallest in magnitude, without  the eigenvector unchanged.  We find that
N=200 is more than sufficient to find $s=s(f)$ for $f\le 10^4$ with no apparent loss in numerical precision.   Figure \ref{f5} shows the amplitudes of the eigenvector components for $f= 10^4$. Even at this large value  the component with maximum amplitude is only $v_6$. 

\begin{thebibliography}{9}
\expandafter\ifx\csname natexlab\endcsname\relax\def\natexlab#1{#1}\fi
\expandafter\ifx\csname bibnamefont\endcsname\relax
  \def\bibnamefont#1{#1}\fi
\expandafter\ifx\csname bibfnamefont\endcsname\relax
  \def\bibfnamefont#1{#1}\fi
\expandafter\ifx\csname citenamefont\endcsname\relax
  \def\citenamefont#1{#1}\fi
\expandafter\ifx\csname url\endcsname\relax
  \def\url#1{\texttt{#1}}\fi
\expandafter\ifx\csname urlprefix\endcsname\relax\def\urlprefix{URL }\fi
\providecommand{\bibinfo}[2]{#2}
\providecommand{\eprint}[2][]{\url{#2}}

\bibitem[{\citenamefont{Doi and Edwards}(1986)}]{Doi86}
\bibinfo{author}{\bibfnamefont{M.}~\bibnamefont{Doi}} \bibnamefont{and}
  \bibinfo{author}{\bibfnamefont{S.~F.} \bibnamefont{Edwards}},
  \emph{\bibinfo{title}{The Theory of Polymer Dynamics;}}
  (\bibinfo{publisher}{Cambridge University Press: 1986},
  \bibinfo{year}{1986}).

\bibitem[{\citenamefont{Flory}(1969)}]{Flory69}
\bibinfo{author}{\bibfnamefont{P.~J.} \bibnamefont{Flory}},
  \emph{\bibinfo{title}{Statistical Mechanics of Chain Molecules}}
  (\bibinfo{publisher}{Butterworth-Heinemann}, \bibinfo{year}{1969}).

\bibitem[{\citenamefont{Ogden et~al.}(2006)\citenamefont{Ogden, Saccomandi, and
  Sgura}}]{Ogden06}
\bibinfo{author}{\bibfnamefont{R.~W.} \bibnamefont{Ogden}},
  \bibinfo{author}{\bibfnamefont{G.}~\bibnamefont{Saccomandi}},
  \bibnamefont{and} \bibinfo{author}{\bibfnamefont{I.}~\bibnamefont{Sgura}},
  \bibinfo{journal}{Proc. R. Soc. A} \textbf{\bibinfo{volume}{462}},
  \bibinfo{pages}{749} (\bibinfo{year}{2006}).

\bibitem[{\citenamefont{Bustamante et~al.}(2000)\citenamefont{Bustamante,
  Smith, Liphardt, and Smith}}]{Bustamante00}
\bibinfo{author}{\bibfnamefont{C.}~\bibnamefont{Bustamante}},
  \bibinfo{author}{\bibfnamefont{S.~B.} \bibnamefont{Smith}},
  \bibinfo{author}{\bibfnamefont{J.}~\bibnamefont{Liphardt}}, \bibnamefont{and}
  \bibinfo{author}{\bibfnamefont{D.}~\bibnamefont{Smith}},
  \bibinfo{journal}{Cur. Opinion Struct. Biol.} \textbf{\bibinfo{volume}{10}},
  \bibinfo{pages}{279} (\bibinfo{year}{2000}).

\bibitem[{\citenamefont{Marko and Siggia}(1995)}]{Marko95}
\bibinfo{author}{\bibfnamefont{J.~F.} \bibnamefont{Marko}} \bibnamefont{and}
  \bibinfo{author}{\bibfnamefont{E.~D.} \bibnamefont{Siggia}},
  \bibinfo{journal}{Macromolecules} \textbf{\bibinfo{volume}{28}},
  \bibinfo{pages}{8759} (\bibinfo{year}{1995}).

\bibitem[{\citenamefont{Bouchiat et~al.}(1999)\citenamefont{Bouchiat, Wang,
  Allemand, Strick, Block, and Croquette}}]{Bouchiat99}
\bibinfo{author}{\bibfnamefont{C.}~\bibnamefont{Bouchiat}},
  \bibinfo{author}{\bibfnamefont{M.~D.} \bibnamefont{Wang}},
  \bibinfo{author}{\bibfnamefont{J.~F.} \bibnamefont{Allemand}},
  \bibinfo{author}{\bibfnamefont{T.}~\bibnamefont{Strick}},
  \bibinfo{author}{\bibfnamefont{S.~M.} \bibnamefont{Block}}, \bibnamefont{and}
  \bibinfo{author}{\bibfnamefont{V.}~\bibnamefont{Croquette}},
  \bibinfo{journal}{Biophys. J.} \textbf{\bibinfo{volume}{76}},
  \bibinfo{pages}{409} (\bibinfo{year}{1999}).

\bibitem[{\citenamefont{Prasad et~al.}(2005)\citenamefont{Prasad, Hori, and
  Kondev}}]{Prasad05}
\bibinfo{author}{\bibfnamefont{A.}~\bibnamefont{Prasad}},
  \bibinfo{author}{\bibfnamefont{Y.}~\bibnamefont{Hori}}, \bibnamefont{and}
  \bibinfo{author}{\bibfnamefont{J.}~\bibnamefont{Kondev}},
  \bibinfo{journal}{Phys. Rev. E} \textbf{\bibinfo{volume}{72}}
  (\bibinfo{year}{2005}).

\bibitem[{\citenamefont{Ogden et~al.}(2007)\citenamefont{Ogden, Saccomandi, and
  Sgura}}]{Ogden07}
\bibinfo{author}{\bibfnamefont{R.~W.} \bibnamefont{Ogden}},
  \bibinfo{author}{\bibfnamefont{G.}~\bibnamefont{Saccomandi}},
  \bibnamefont{and} \bibinfo{author}{\bibfnamefont{I.}~\bibnamefont{Sgura}},
  \bibinfo{journal}{Comp. Math. Appl.} \textbf{\bibinfo{volume}{53}},
  \bibinfo{pages}{276} (\bibinfo{year}{2007}).

\bibitem[{\citenamefont{Abramowitz and Stegun}(1974)}]{Abramowitz74}
\bibinfo{author}{\bibfnamefont{M.}~\bibnamefont{Abramowitz}} \bibnamefont{and}
  \bibinfo{author}{\bibfnamefont{I.}~\bibnamefont{Stegun}},
  \emph{\bibinfo{title}{Handbook of Mathematical Functions with Formulas,
  Graphs, and Mathematical Tables}} (\bibinfo{publisher}{Dover, New York},
  \bibinfo{year}{1974}).

\end{thebibliography}


\end{document}